\documentclass[desactivate]{aa}  

\usepackage{graphicx}
\usepackage{txfonts}
\usepackage{hyperref}
\usepackage{natbib}
\usepackage{xcolor}
\usepackage{makecell}

\bibpunct{(}{)}{;}{a}{}{,}

\newcommand{\boldhat}[1]{\ensuremath\boldsymbol{\hat{#1}}}

\begin{document}


\title{Unravelling sub-stellar magnetospheres}
\subtitle{}

\author{Robert D. Kavanagh\inst{1,2}, Harish K. Vedantham\inst{1,3}, Kovi Rose\inst{4,5}, Sanne Bloot\inst{1, 3}}

\institute{ASTRON, The Netherlands Institute for Radio Astronomy, Oude Hoogeveensedijk 4, 7991 PD Dwingeloo, The Netherlands \\\email{kavanagh@astron.nl}
\and
Anton Pannekoek Institute for Astronomy, University of Amsterdam, 1098 XH Amsterdam, The Netherlands
\and
Kapteyn Astronomical Institute, University of Groningen, PO Box 72, 97200 AB Groningen, The Netherlands
\and
Sydney Institute for Astronomy, School of Physics, The University of Sydney, NSW 2006, Australia
\and
CSIRO Astronomy and Space Science, PO Box 76, Epping, NSW 1710, Australia}

\date{Received 2 Sep 2024; accepted ...}

\abstract{At the sub-stellar boundary, signatures of magnetic fields begin to manifest at radio wavelengths, analogous to the auroral emission of the magnetised solar system planets. This emission provides a singular avenue for measuring magnetic fields at planetary scales in extrasolar systems. So far, exoplanets have eluded detection at radio wavelengths. However, ultracool dwarfs (UCDs), their higher mass counterparts, have been detected for over two decades in the radio. Given their similar characteristics to massive exoplanets, UCDs are ideal targets to bridge our understanding of magnetic field generation from stars to planets. In this work, we develop a new tomographic technique for inverting both the viewing angle and large-scale magnetic field structure of UCDs from observations of coherent radio bursts. We apply our methodology to the nearby T8 dwarf WISE J062309.94-045624.6 (J0623) which was recently detected at radio wavelengths, and show that it is likely viewed pole-on. We also find that J0623's rotation and magnetic axes are misaligned significantly, reminiscent of Uranus and Neptune, and show that it may be undergoing a magnetic cycle with a period exceeding 6 months in duration. These findings demonstrate that our method is a robust new tool for studying magnetic fields on planetary-mass objects. With the advent of next-generation low-frequency radio facilities, the methods presented here could facilitate the characterisation of exoplanetary magnetospheres for the first time.}


\authorrunning{Kavanagh et al.}

\maketitle


\section{Introduction}

Magnetic fields play a key role in regulating the habitable conditions of exoplanets. The strength and geometry of planetary magnetospheres are thought to play a fundamental role as to how their atmospheres are impacted by both plasma outflows from their host stars \citep{owen14, carolan21b} and energetic particles \citep{herbst19, rodgers-lee23}. Measuring magnetic fields on planets also provides a unique probe of their interiors \citep[e.g.][]{rodriguez-mozos22}, and their presence can facilitate the detection of orbiting companions via radio observations \citep{kavanagh23}. With the ever-increasing number of newly-discovered extrasolar worlds, so too does our motivation to characterise their magnetic fields.

To date, measurement of magnetic fields on a stellar scale has been facilitated by observable signatures of the Zeeman effect in spectral lines, which can facilitate the reconstruction of surface magnetic fields maps via the tomographic Zeeman-Doppler imaging (ZDI) technique \citep{kochukhov21}. However, it becomes unfeasible to measure these signatures at the star-planet boundary. The late M-dwarfs and brown dwarfs that straddle this boundary, collectively known as ultracool dwarfs (UCDs), are intrinsically faint. Additionally, their spectra are dominated by molecular bands, which are difficult to model \citep{kuzmychov17, behmard19}. Therefore, measuring magnetic fields on these objects via the Zeeman effect is not feasible with current technology \citep[although see][]{berdyugina17}. The prospects for measuring the Zeeman effect on an exoplanet are even less promising given the contrast with their host stars. However, a distinct sign of magnetic fields begins to manifest at the sub-stellar boundary. 

Prior to the early 2000s, there was little evidence of magnetic fields on UCDs. This abruptly changed however with the first reported detection of radio bursts from the M9 dwarf LP~944-20 by \citet{berger01}. These bursts were interpreted as incoherent gyrosynchrotron emission, requiring the presence of a magnetic field with a strength of $\sim5$~G. Later on, \citet{hallinan08} demonstrated that radio bursts from two UCDs of spectral types M8.5 and L3.5 are likely driven by a coherent mechanism, the \textit{electron cyclotron maser} (ECM) instability. This again implied the presence of magnetic fields, however with a strength of the order of 1~kG.

These early signs of magnetic fields on UCDs provided the opportunity to study magnetic field generation from stellar to planetary scales for the first time. By considering the internal heat flux within the convective zones, \citet{christensen09} derived a scaling law which successfully predicted the field strengths from the solar system planets up to early G-type main-sequence stars. Given their rapid rotation, this scaling law predicts dipole-like magnetic fields with strengths of the order of 1~kG for UCDs, in line with those estimated from their radio emission. This consistency further strengthened the validity of this scaling law, which in turn means that it can be utilised to inform us about the field strengths of exoplanets. However, it cannot inform us about the geometry of magnetic fields on UCDs, which is both important for detecting satellites orbiting within magnetospheres \citep{kavanagh23} and the retention of an atmosphere (see above).

To date, 34 UCDs have been detected in the radio \citep[see][]{vedantham20c, kao22, tang22, vedantham23, rose23, magaudda24}. In general, they show quiescent emission that is weakly modulated with a  low polarisation fraction, on top of which bright circularly-polarised bursts are often observed \citep{williams18}. The common consensus for the quiescent emission of UCDs is that energetic electrons are trapped within their large-scale magnetospheres, producing synchrotron emission \citep{hallinan06, kao24}. This idea has been further strengthened with the recent reports of resolved radio emission around the M8.5 UCD LSR~J1835+3259, which bears a striking resemblance to the radiation belts seen on the Earth and Jupiter \citep{kao23, climent23}.

On the other hand, the bright circularly-polarised bursts sometimes observed from UCDs are thought to be driven by the ECM instability \citep{hallinan07, hallinan15, kao22}. The conditions necessary to produce ECM emission are complex, and are described in detail by \citep{treumann06}. In short, a source of energetic electrons must be injected into the magnetosphere. As these charges accelerate towards higher magnetic latitudes, those with small pitch angles (the angle between their velocity vector and the magnetic field) are lost to the atmosphere. However, those with large pitch angles undergo a magnetic mirroring effect, wherein they reflect back along the field line in the direction from which they came. This results in a velocity distribution of accelerated electrons that resembles a horseshoe in velocity space, meaning that only electrons with large pitch angles remain spiralling in the field. Electrons with this distribution can then amplify background radio emission exponentially, resulting in bright coherent radio bursts.

ECM emission occurs at harmonics of the cyclotron frequency $\nu_\text{c}$, which scales linearly with the magnetic field strength \citep{dulk85}. However, ECM emission that is observable generally implies it occurs at either the fundamental or first harmonic of $\nu_\text{c}$ \citep{melrose82}. Therefore, ECM emission is a \textit{direct measure} of magnetic field strength. It is also a much more robust tracer of field strength compared to synchrotron emission, which is inherently broadband due to relativistic effects, making the determination of the B field partially degenerate with the relativistic Lorentz factor of the charges. However, the ability for synchrotron to produce resolvable structure at radio wavelengths (as described for LSR J1835+3259 above) will likely greatly complement ECM-oriented methods for studying magnetic field strengths in the near future.

Another defining characteristic of ECM emission is that its amplification process is very sensitive to the angle the background radiation makes with the magnetic field. As a result, amplification only occurs at specific angles relative to the magnetic field. This produces an anisotropic beam pattern that resembles a hollow cone centered on the magnetic field line. Since the emission is only visible when the magnetic field is oriented correctly from an observer's perspective, ECM emission provides a unique probe of the underlying magnetic field geometry \citep[e.g.][]{lynch15, bastian22, kavanagh23, bloot24}.

In the solar system, the ECM instability is responsible for the generation of bright emission from all of the magnetised planets near their magnetic poles. This is either due to sub-Alfv\'enic interactions with orbiting satellites, the deposition of energy carried by the solar wind onto their magnetospheres, or shearing between the magnetosphere and the surrounding plasma environment \citep{zarka07}. This emission is often referred to as being `auroral'. Despite extensive searches, there has yet to be a conclusive detection of radio emission from an exoplanet \citep[see][]{turner21, turner24}. As a result, their magnetic field strengths remain unknown to date. Assuming their field strengths are comparable to those of the solar system planets, sensitive ultra-low frequency radio facilities such as GO-LoW \citep{knapp24}, LOFAR2.0 \cite{edler21}, and SKA-Low \citep{braun17} will likely be necessary for their direct detection.

Despite the challenges in detecting exoplanetary magnetic fields, radio-emitting UCDs indicative of ECM emission are perfect targets for bridging our understanding of magnetism from stars to planets. UCDs are massive analogues for giant exoplanets. Understanding the generation, diversity, and evolution of magnetic fields on UCDs can therefore inform us about the same processes at planetary scales, as their radio emission is a tracer of magnetic fields. Then with the advent of next-generation radio telescopes, the same methodologies for studying UCD magnetism will likely become applicable to exoplanets.

To this end, in this work we present a new framework for retrieving the large-scale magnetic field geometry of UCDs from their radio morphology. This method is effectively a tomographic technique analogous to the ZDI method, relying on the time evolution of the emission cone visibility that is characteristic to ECM emission. We demonstrate its effectiveness by applying it to a recently-detected UCD with an unusual radio lightcurve.


\section{Puzzling periodic pulses}

\citet{rose23} recently reported the detection of circularly-polarised radio bursts at around 1 to 1.5~GHz from the T8 dwarf WISE~J062309.94-045624.6 (hereafter J0623). The bursts repeat with a period of 1.9~hours, comparable to the rotation periods inferred for T~dwarfs from their photometric variability \citep{tannock21}. The high brightness temperature and circular polarisation fraction of the emission implies that a coherent emission mechanism is at play in its magnetosphere \citep{melrose82}.

Circularly-polarised radio bursts from UCDs typically manifest as narrow pulses \citep[e.g.]{hallinan08, kao18}. As a result, the information that can be extracted from them is limited. The morphology of those from J0623 on the other hand show a complex structure. Over 1.9 hours, the MeerKAT observations from \citet{rose23} first show a bright burst that reaches a level of around 3-4~mJy, lasting around 15 minutes. About 10 minutes later, at least 2 or 3 bursts are then seen consecutively over the next 50 minutes, each peaking at around 2~mJy. All of the bursts are polarised in a left-handed sense. This pattern then repeats twice more. 

The question is, what is the origin of this repeating pattern? As the emission is coherent, the emission could be produced either via plasma emission or the ECM instability \citep{dulk85}. However, the densities required to drive plasma emission are likely orders of magnitude larger than those present on brown dwarfs given the emission frequency \citep{richey-yowell20}. Furthermore, it is difficult to reconcile the repeating pattern seen on J0623 with stochastic flaring. This may instead imply that a stable supply of energy is provided to the magnetosphere.

The alternative emission scenario for J0623 is ECM, as has been inferred for many UCDs \citep[e.g.][]{hallinan07, hallinan15, kao22}. Given its interesting pulse structure, this object may provide the opportunity to extract geometric information from underlying magnetic field by exploiting the beamed property of ECM emission. We recently took such an approach to infer the large-scale magnetic field of the mid M-dwarf AU Mic \citep{bloot24}. Analysing over 250 hours of radio data on the star, we found that AU Mic exhibits a repeated structure indicative of emission cones sweeping across the line of sight in each rotation phase. A similar phenomenon is seen on Jupiter \citep{marques17}, which exhibits an auroral ring at its magnetic poles. By constructing a simple geometric model for the visibility of emission cones in an auroral ring configuration, we were able to reproduce the overall shape of the radio lightcurve of AU Mic. Our inferred geometry and field strength was also consistent with those inferred via ZDI during the same epoch by \citet{donati23}, highlighting the promise of our approach as an alternative pathway for extracting magnetospheric structure.

However, the auroral ring approach fails to reproduce the observed lightcurve of J0623, in that it only predicts symmetric lightcurves. Instead, we turn towards the idea of \textit{active field lines} (AFLs). This concept has previously been explored to interpret radio signatures from mid M-dwarfs to early L-dwarfs \citep{lynch15, bastian22}.


\section{Unravelling the magnetosphere of J0623}
\label{sec:MeerKAT fitting}

In the AFL paradigm, a subset of field lines are chosen to emit ECM within a large-scale dipolar magnetic field. The field is characterised by its magnetic obliquity $\beta$, the angle between the magnetic and rotation axes. The rotation axis itself is inclined relative to the line of sight by the angle $i$ (see \ref{sec:coordinates} for more details). Each AFL is located by its longitude $\phi_\text{B}$ in the magnetic equator. Given that dipolar magnetic field lines are symmetric, emission at a frequency $\nu$ occurring at a magnetic co-latitude $\theta_\text{B}$ can also occur at the co-latitude $\pi - \theta_\text{B}$ with the same frequency.

The visibility of ECM emission from AFLs depends on their orientation with respect to the line of sight. Each emission cone opens outwards from the field line with the angle $\alpha$ away from the surface, with a thickness of $\Delta\alpha$. The emission is visible when the angle $\gamma$ measured between the line of sight and the magnetic field vector that each cone aligned with is in the range $\alpha\pm\Delta\alpha/2$. Due to both the rotation and magnetic precession, the angle each cone forms with the line of sight over time is complex. We assume that when the alignment is perfect (i.e. $\gamma = \alpha$), the observer sees a flux density $F_\text{B}$. 

We parameterise the decay in flux density observed from each emission cone as it becomes misaligned via the following equation:
\begin{equation}
F = F_\text{B} \bigg[ \exp \bigg\{- \frac{1}{2} \Big(\frac{\gamma_\text{N} - \alpha}{\Delta\alpha}\Big)^2 \bigg\} - \exp \bigg\{- \frac{1}{2} \Big(\frac{\gamma_\text{S} - \alpha}{\Delta\alpha}\Big)^2 \bigg\} \bigg] .
\label{eq:line flux}
\end{equation}
Note that here $\gamma_\text{N}$ and $\gamma_\text{S}$ refer to the angles the cones in the Northern and Southern magnetic hemisphere form with the line of sight. Our method for computing these `beam' angles for a given set of the parameters introduced above is described in Appendix~\ref{sec:beam angle}. In the case that there are multiple AFLs, we sum Equation~\ref{eq:line flux} for each line. Assuming emission occurs via the same magnetoionic mode in both hemispheres, the flux density received from the two cones will have opposite circular polarisations. This implies that flux cancellation is possible if both cones are visible simultaneously. Here we implicitly assume that emission from the Northern hemisphere has a positive Stokes V flux density.

The question now is: what set of parameters reproduce the observed radio lightcurve of J0623, and how many AFL are needed? To answer this, we need to explore the likelihood space defined by Equation~\ref{eq:line flux} given the observed data. For this, we converged on using UltraNest\footnote{\url{https://johannesbuchner.github.io/UltraNest/}} \citep{buchner21a}, which utilises the nested sampling Monte Carlo algorithm MLFriends \citep{buchner16, buchner19}. Since there are slight differences in the MeerKAT lightcurves of J0623 from 950 to 1150~MHz and 1300 to 1500~MHz, we construct a joint fit for the two bands (see Appendix~\ref{sec:MeerKAT data prep} for details). Our sampling method in also described further in Appendix~\ref{sec:sampling}.

\begin{figure}
\centering
\includegraphics[width=\linewidth]{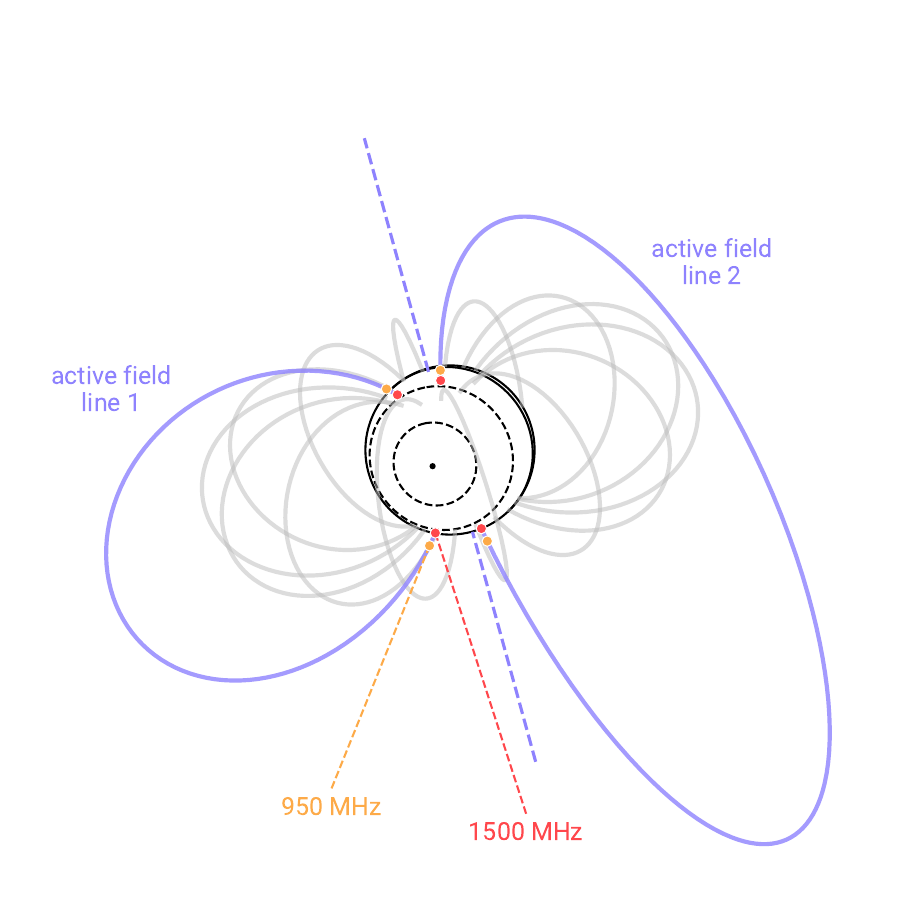}
\caption{The magnetic field geometry inferred for the T8 dwarf J0623. The black dot on the surface shows the rotation axis, which the observer effectively sees pole-on. The dashed lines on the surface are lines of constant latitude in $30\degr$ intervals, with the solid line showing the equator. Dipolar field lines with a size of 3 radii are shown in grey, and the two active field lines which emit electron cyclotron maser emission are shown in purple. The emission points at 950 and 1500 MHz are marked on each active field line. The brown dwarf has a large magnetic obliquity, reminiscent of Uranus and Neptune. The magnetic axis is also shown as a dashed purple line intersecting the brown dwarf. Note that the active field lines are shown here with their minimum size (see Section~\ref{sec:field strength}).}
\label{fig:field config}
\end{figure}

\begin{figure*}
\centering
\includegraphics[width=\textwidth]{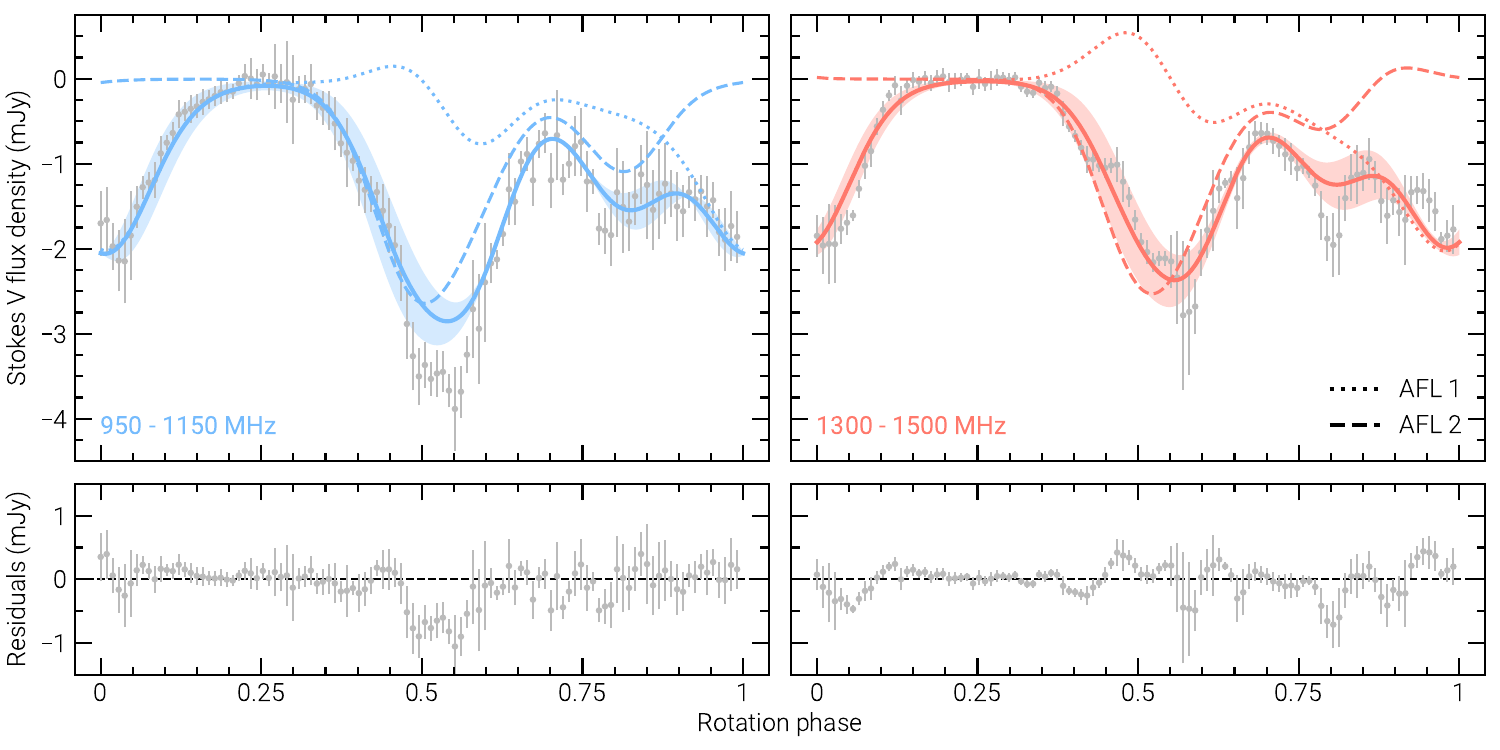}
\caption{The best fit to the MeerKAT data of J0623 from 950 to 1150 MHz (top left) and 1300 to 1500 MHz (top right). $1\sigma$ error bars are shown for each data point. The solid curves and shaded regions show the average and $3\sigma$ variance of the model lightcurve over each band. The dotted and dashed curves illustrate the averaged individual contributions of each active field line to the lightcurve. The lower panels show the residuals of the data with the average lightcurve (observed minus model flux density).}
\label{fig:MeerKAT lightcurves}
\end{figure*}

We sample the likelihood space by considering scenarios with 1 and 2 AFLs. Both produce similar geometric configurations for the rotation and magnetic axes, however the Bayesian evidence $\log\mathcal{Z}$ provided by UltraNest is significantly higher for the scenario with 2 AFLs. The difference in $\log\mathcal{Z}$ from 1 to 2 AFLs is $\sim1800$, which strongly favours 2 over 1 AFL according to the Jeffreys scale \citep[e.g.][]{callingham15}. The geometric configuration inferred in the 2 AFL scenario is shown in Figure~\ref{fig:field config}, and the confidence intervals obtained for the system parameters both scenarios are listed in Table~\ref{table:MeerKAT values}. In Figure~\ref{fig:MeerKAT lightcurves}, we compare the maximum likelihood solution to the MeerKAT data of J0623.

\begin{figure}
\centering
\includegraphics[width=\linewidth]{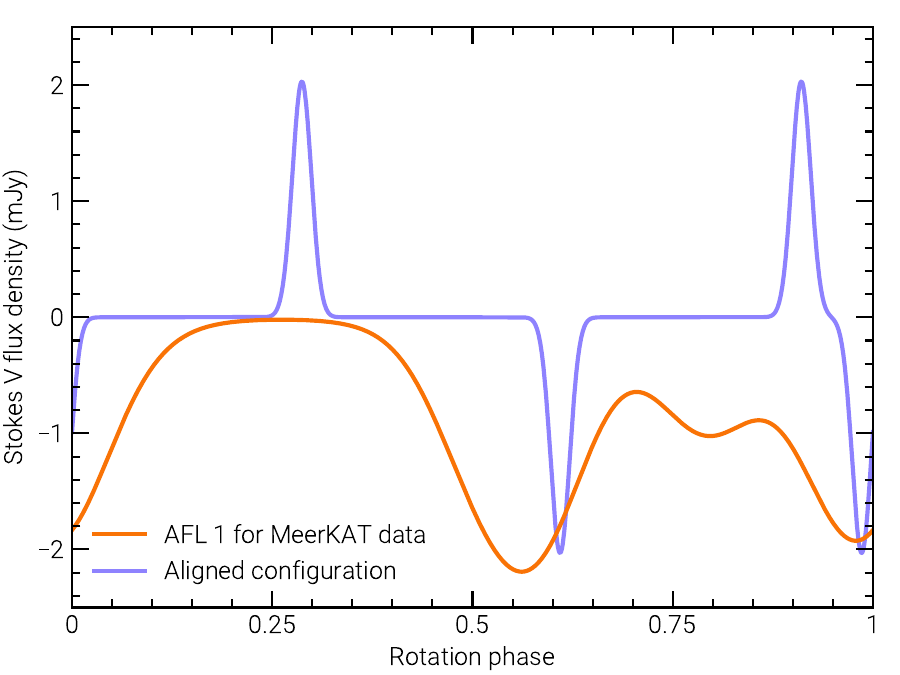}
\caption{Comparison of the lightcurve of AFL 1 inferred from the MeerKAT data (orange) versus the scenario where J0623 is viewed equator-on with a magnetic obliquity of zero in an `aligned' configuration (purple). In the aligned scenario, the visibility of the emission decreases significantly. Our inferred geometry therefore may hint at a detection bias at radio wavelengths for brown dwarfs in pole-on configurations with large magnetic obliquities.}
\label{fig:aligned comp}
\end{figure}

We infer that J0623 is viewed near pole-on, and has a large magnetic obliquity of $\sim80\degr$. This is interesting, in that \citet{kavanagh23} demonstrated that blind radio surveys are biased towards finding signatures of star-planet interactions from stars with the same geometry, assuming large cone opening angles. This geometry results in high duty cycles (the fractional amount of time the signal is visible), which for J0623 is around 70\% \citep{rose23}. Given that J0623 was detected blindly in the Rapid ASKAP Continuum Survey \citep[RACS;][]{mcconnell20}, our results may imply that the same detection bias is also valid for ultracool dwarfs at radio wavelengths. To further illustrate this, in Figure~\ref{fig:aligned comp}, we compare the lightcurve of AFL 1 inferred from the MeerKAT data to a system with the same parameters, except with an inclination of $90\degr$ and a magnetic obliquity of $0\degr$. We see that the visibility of the AFL drops significantly in the aligned case. The duty cycle of emission above 0.1 mJy in the inferred case is 80\%, whereas for the aligned case it is 23\%.

The large magnetic obliquity we find for J0623 is also of interest, in that it is reminiscent to that of Uranus and Neptune \citep{bagenal13}. This may provide some insight into both the composition and dynamo generation of the interior \citep{kao16, kimura21}. Due to its obliquity, the magnetic poles of J0623 effectively precess in the equatorial plane. On Neptune, in-situ observations have shown that the exposure of the magnetic poles to the equatorial region produces favourable conditions for magnetic reconnection, driven by the impact of the solar wind \citep{jasinski22}. Magnetic reconnection itself is known to produce ECM emission in the solar system \citep{cowley01}, and is thought to be a viable option for driving coherent radio bursts on brown dwarfs and stars \citep[e.g.][]{nichols12, das21, shultz22}. We discuss the possible driving mechanisms of the coherent emission from J0623 further in Section~\ref{sec:plasma}.

\begin{table}
\renewcommand{\arraystretch}{1.6}
\caption{The inferred properties obtained for J0623 assuming 1 or 2 AFLs emit in its large-scale magnetosphere. Each parameter is described in Section~\ref{sec:MeerKAT fitting}, except for $\phi_0$ which is the rotation phase of J0623 at the start of the MeerKAT observations. The nominal values shown are the 50th percentile, and the lower and upper values are the 16th and 84th percentiles. Values quoted without errors have errors less than the least significant digit. The subscripts 1 and 2 refer to each AFL. At the bottom of the table, we show the Bayesian evidence and maximum likelihood that UltraNest converges on. Note that the co-latitude values listed correspond to emission at the maximum frequency of 1500~MHz (see Appendix~\ref{sec:model frequency}). We also provide the reduced $\chi^2$ values, which we compute from the data with the average lightcurve over each band. The low and high subscripts refer to the two bands of the MeerKAT data, which are 950 to 1150 MHz and 1300 to 1500 MHz. The posterior distributions for the parameters inferred for the 2 AFL scenario is shown in Figure~\ref{fig:MeerKAT corner plot}. Note that some of the parameters listed are degenerate (see Appendix~\ref{sec:degeneracies}).}
\label{table:MeerKAT values}
\centering
\begin{tabular}{c|cc}
Parameter & 1 AFL & 2 AFLs \\
\hline
$\cos i$ & $0.95$ & $0.99$ \\
$\beta$ ($\degr$) & $57.99_{-0.40}^{+0.36}$ & $80.85_{-0.24}^{+0.25}$ \\
$\phi_0$ & $0.59$ & $0.80$ \\
$\alpha$ ($\degr$) & $89.92_{-0.14}^{+0.06}$ & $69.82_{-0.33}^{+0.32}$ \\
$\Delta\alpha$ ($\degr$) & $9.97_{-0.05}^{+0.02}$ & $3.87_{-0.11}^{+0.12}$ \\
$\theta_1$ ($\degr$) & $33.19_{-0.25}^{+0.26}$ & $27.57\pm0.20$ \\
$\theta_2$ ($\degr$) & $-$ & $18.25_{-0.17}^{+0.18}$ \\
$\phi_1$ ($\degr$) & $247.12\pm0.23$ & $127.42_{-0.47}^{+0.48}$ \\
$\phi_2$ ($\degr$) & $-$ & $215.45_{-0.65}^{+0.68}$ \\
$F_1$ (mJy) & $2.40\pm0.01$ & $2.04\pm0.02$ \\ 
$F_2$ (mJy) & $-$ & $2.67\pm0.05$ \\
\hline
$\log\mathcal{Z}$ & $-2314$ & $-511$ \\
$\log\mathcal{L}$ & $-2265$ & $-433$ \\
${\chi^2}_\text{red,low}$ & 4.15 & 1.33 \\
${\chi^2}_\text{red,high}$ & 3.94 & 2.14 \\
\end{tabular}
\end{table}



\section{Potential magnetic field evolution}
\label{sec:cycle}

J0623 was also observed with the Australian Telescope Compact Array (ATCA) for 6 hours on 14 Aug 2022 and for 11 hours on 16 Dec 2022 \citep{rose23}. While these observations predate the MeerKAT observations which were taken in Mar 2023, they have a lower signal to noise ratio in Stokes V. This was our main motivation to primarily analyse the MeerKAT data over the ATCA data. However, it is interesting to explore how our predicted field geometry for J0623 holds up at these other epochs. Particularly, does the magnetic field or positions of the active field lines change significantly?

Firstly, we reprocessed the ATCA data presented by \citet{rose23} to produce the Stokes V lightcurves at different frequency bands. This is described in more detail in Appendix~\ref{sec:ATCA data reduction}. For the Aug data, we obtained two lightcurves from 1.3 to 1.77 GHz and 1.77 to 2.43 GHz. From 2.43 to 3.1 GHz, there is no signal visible. We therefore use this upper band to estimate the noise, which we measure to be 0.34 mJy. Similarly for the Dec data, we created lightcurves from 1.3 to 1.6 GHz and 1.6 to 2.1 GHz. The band from 2.1 to 3.1 GHz shows no signal, from which we extract a noise value of 0.33 mJy.

\begin{figure}
\centering
\includegraphics[width = \linewidth]{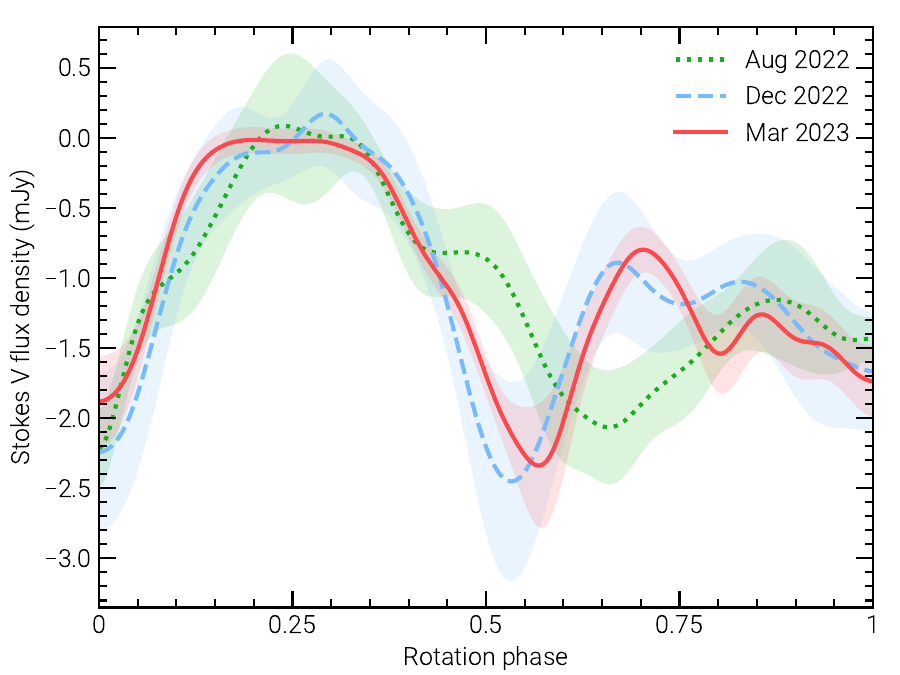}
\caption{Comparison of the Stokes V lightcurves of J0623 on obtained with ATCA on 14 Aug 2022 ($1.3-1.77$~GHz) and 16 Dec 2022 ($1.3-1.6$~GHz), and with MeerKAT on 27 Mar 2023 ($1.3-1.5$~GHz). The data have been linearly interpolated for 500 points uniformly spaced in rotation phase and smoothed with a Gaussian with a width of $5\sigma$ using SciPy's \texttt{gaussian\_filter1d} function. The shaded regions show the $1\sigma$ variance. The structure appears to change over a timescale of around 6 months, which we speculate could be due to a magnetic cycle (see Section~\ref{sec:cycle}).}
\label{fig:lightcurve comp}
\end{figure}

The ATCA data show similar structure overall to the MeerKAT data, but some differences become noticeable for the Aug 14 data in particular. We compare the MeerKAT and ATCA lightcurves in Figure~\ref{fig:lightcurve comp}. To explore if this change may pertain to a change in the underlying physical parameters modelled here, we re-apply the same sampling method as for the MeerKAT data. We first allowed only the 2 AFL parameters to vary, fixing the other parameters to those that give maximum likelihood for the MeerKAT data. We then ran an additional test for each epoch allowing the magnetic obliquity to also vary. The results are listed in Table~\ref{table:ATCA values}. We also show the fits to the data in Figure~\ref{fig:ATCA lightcurves}. For the Dec data, which were taken closer to the MeerKAT data, we find that the Bayesian evidence is higher if the magnetic obliquity remains unchanged from that inferred for the MeerKAT data. The values inferred also imply a slight change in parameters corresponding to the second AFL.

For the Aug data on the other hand, the evidence favours a magnetic obliquity of $\sim50\degr$. The increase in $\log\mathcal{Z}$ compared to the case where the other parameters are kept fixed is $\sim6$. However, the reduced $\chi^2$ for the fits become worse for the noise and low frequency band when compared to the case where $\beta$ is fixed as the value inferred from the MeerKAT data. The favoured results here also imply changes in the AFL properties. In particular, they imply that the 2nd AFL has drifed by about 20$\degr$ in longitude. 

These results may imply that the magnetic axis has drifted by $\sim30\degr$ over the course over the course of 6 months. Obtaining follow-up data of J0623 with MeerKAT is worth pursuing to explore this further, in that it could allow us to track a potential magnetic cycle on a brown dwarf for the first time. So far, evidence of magnetic cycles past the sub-stellar boundary have only been loosely inferred via changes in magnetic polarity in radio bursts \citep{route16}. Additionally, tracking potential drifts in the positions of the AFLs may allude to the orbital motion of a satellite within the magnetosphere. However, this is purely speculative at this time.

\begin{table*}
\renewcommand{\arraystretch}{1.6}
\caption{The same as Table~\ref{table:MeerKAT values}, except now the values are estimated from the ATCA data of J0623 obtained in Aug and Dec of 2022. Results are shown for allowing all active field line parameters to vary, keeping all the remaining parameters fixed, and then also allowing the magnetic obliquity $\beta$ to also vary. The values of $\theta_1$ and $\theta_2$ correspond to emission at 2.43~GHz for Aug and 2.1~GHz for Dec respectively (see Appendix~\ref{sec:model frequency}). The reduced $\chi^2$ values are again computed between the data and the average model lightcurve over the low and high frequency bands. The low band is from 1.3 to 2.43 GHz for the Aug data, and 1.3 to 2.1 GHz for Dec data. We also include the reduced $\chi^2$ between the noise bands, which are 2.43 to 3.1 GHz and 2.1 to 3.1 GHz for the Aug and Dec data respectively.}
\label{table:ATCA values}
\centering
\begin{tabular}{c|cc|cc}
Parameter & \multicolumn{2}{c|}{14/8/2022} & \multicolumn{2}{c}{16/12/2022} \\
\hline
$\beta$ ($\degr$) & Fixed & $50.48_{-0.79}^{+0.78}$ & Fixed & $50.81_{-1.05}^{+1.00}$ \\
$\theta_1$ ($\degr$) & $25.01\pm0.20$ & $42.35_{-0.18}^{+0.17}$ & $23.98_{-0.30}^{+0.29}$ & $42.90_{-0.26}^{+0.27}$ \\
$\theta_2$ ($\degr$) & $21.11_{-0.22}^{+0.21}$ & $39.62_{-0.33}^{+0.32}$ & $21.22_{-0.39}^{+0.37}$ & $38.97_{-0.46}^{+0.42}$ \\
$\phi_1$ ($\degr$) & $125.54\pm0.36$ & $124.14_{-0.49}^{+0.55}$ & $127.74_{-0.57}^{+0.58}$ & $123.81_{-0.63}^{+0.67}$ \\
$\phi_2$ ($\degr$) & $227.63_{-0.49}^{+0.47}$ & $235.59_{-0.65}^{+0.63}$ & $227.52_{-0.84}^{+0.86}$ & $237.19_{-0.67}^{+0.64}$ \\
$F_1$ (mJy) & $2.34_{-0.06}^{+0.07}$ & $1.46_{-0.03}^{+0.02}$ & $2.05\pm0.09$ & $1.05\pm0.03$ \\
$F_2$ (mJy) & $2.52\pm0.07$ & $1.21_{-0.09}^{+0.10}$ & $2.14\pm0.09$ & $2.92_{-0.55}^{+0.67}$ \\
\hline
$\log\mathcal{Z}$ & $-577$ & $-571$ & $-282$ & $-308$ \\
$\log\mathcal{L}$ & $-544$ & $-535$ & $-251$ & $-276$ \\
${\chi^2}_\text{red,low}$ & 3.65 & 2.14 & 1.41 & 1.75 \\
${\chi^2}_\text{red,high}$ & 1.41 & 2.39 & 1.80 & 0.94 \\
${\chi^2}_\text{red,noise}$ & 0.55 & 0.85 & 0.71 & 0.41 \\
\end{tabular}
\end{table*}


\section{The magnetic field strength of J0623}
\subsection{Constraints from radio modelling}
\label{sec:field strength}

It is important to note that our inversion method does not depend on the field strength at the magnetic poles $B_0$ (see Appendix~\ref{sec:model frequency}). Instead, it depends on the quantity $B_0/L^3$, where $L$ is the maximum extent of the AFL in the magnetic equator or `loop size'. Equation~\ref{eq:constant} relates $B_0/L^3$ to the co-latitude of the emission point $\theta_\text{B}$ and the frequency of emission $\nu$. While we cannot break the degeneracy between $B_0$ and $L$, we can still place constraints on their values.

The first constraint we can place is on the minimum size of $L$. For emission to be visible, the maximum frequency has to occur above the point where the AFL touches the surface. Via Equation~\ref{eq:field coordinates}, this means that
\begin{equation}
\frac{L}{R} > \frac{1}{\sin^2\theta_\text{B}},
\end{equation}
where $R$ is the radius. Since we obtain the co-latitude of the emission point on each co-latitude separately, each AFL has its own minimum size. For the favoured 2 AFL scenario inferred from the MeerKAT data (Table~\ref{table:MeerKAT values}), the minimum loop sizes of AFLs 1 and 2 are 4.67 and 10.2 $R$ (as shown in Figure~\ref{fig:field config}). Similarly, for our fits to the 2 ATCA epochs, we estimate minimum loop sizes for AFLs 1 and 2 of 2.37 and 2.65~$R$ in Aug, and 6.51 and 8.21~$R$ in Dec.

The minimum loop sizes also allow us to place a lower limit on the magnetic field strength. This is an important quantity to determine, in that it can inform us about the internal dynamo \citep{kao16}. Given the minimum loop size, one can show via Equation~\ref{eq:constant} that
\begin{equation}
B_0 > \frac{\nu}{1.4(1 + 3\cos^2\theta_\text{B})^{1/2}} .
\end{equation}
Again, both AFLs provide a constraint on this. However, the constraint provided by the larger magnetic co-latitude should be adopted, in that it provides the largest lower limit, which is required for emission from both AFLs to be visible. From the MeerKAT data, we estimate a lower limit of 585~G on the dipolar field strength of J0623. For the ATCA data on the other hand, we estimate a lower limit of 1339~G in Aug and 1177~G in Dec. These are significantly larger than the estimate from the MeerKAT data as we predict emission up to 3.1~GHz, albeit at a level that is generally below the noise in the ATCA data (see Figure~\ref{fig:ATCA lightcurves}). While the higher observing band is likely the primary factor for the larger field strength estimates, an increase in the minimum field strength towards the Aug 2022 epoch may also support the idea for a magnetic cycle as discussed in Section~\ref{sec:cycle}.

The method deployed here provides a more robust constraint on magnetic field strengths compared to previous works, in that it is informed by the magnetic co-latitude of the emission cone. In the absence of this information, one has to assume that the emission comes from the magnetic pole \citep[e.g.][]{kao18}. For instance, lacking this information, our estimated lower limit from the MeerKAT data would instead be 536~G. We also note that we are assuming that the emission occurs at the fundamental of the cyclotron frequency. However, emission at the first harmonic is also possible, in which case our lower limit here should be halved.


\subsection{Comparison to scaling laws}

It is also worth comparing our field strength estimates for J0623 to that predicted by the scaling law presented by \citet{christensen09}. This scaling law gives a prescription for the dipolar field strength as a function of the mass $M$, luminosity $L$, and radius $R$ \citep{reiners09, reiners10}:
\begin{equation}
B_0 = 3.4_{-2.0}^{+2.2} \bigg(\frac{ML^2}{R^7}\bigg)^{1/6} \bigg(1 - \frac{0.17~M_\text{Jup}}{M}\bigg)^3~\text{kG} .
\label{eq:christensen scaling law}
\end{equation}
Note that all quantities here are in solar units, and $M_\text{Jup}$ is the mass of Jupiter.

To estimate the mass, luminosity, and radius of J0623, we use version 2 of the solar metallicity Sonora Diamondback\footnote{\url{https://zenodo.org/records/12735103}} evolutionary models presented by \citet{morley24} for UCDs. These provide the effective temperature $T_\text{eff}$, surface gravity $\log g$, and luminosity as a function of age for a given mass. These quantities vary over time, as UCDs expand and cool down via the burning of deuterium. While \citet{morley24} also present models that include the effects of atmospheric clouds, we choose to use the cloud-free models as \citet{morley24} states that these better-reproduce the measured photometry of T dwarfs.

To obtain the quantities needed for Equation~\ref{eq:christensen scaling law}, we need some input quantity to interpolate with between the different evolutionary tracks. For this, we use the estimated values of $T_\text{eff}$ and $\log g$ for J0623 obtained by \citet{zhang21} from modelling its low-resolution near-infrared spectrum. These are $T_\text{eff} = 743_{-51}^{+53}$~K and $\log g = 4.70_{-0.42}^{+0.47}$. Note that these values are co-variant. For each set of $\log g$ and $T_\text{eff}$, we linearly interpolate between the values for the mass, radius, and luminosity from the evolutionary tracks using SciPy's \texttt{LinearNDInterpolator} function. We note that the samples for $T_\text{eff}$ and $\log g$ from \citet{zhang21} have a 93\% overlap with the evolutionary tracks from \citet{morley24}. Those that lie outside the bounds of the tracks are discarded.

Interpolating between the evolutionary tracks, we obtain the following estimates for the properties of J0623: $M = 22.22_{-10.30}^{+15.53}~M_\text{Jup}$, $R = 1.01\pm0.14~R_\text{Jup}$, and $\log L = -5.55_{-0.19}^{+0.22}~L_\sun$. Plugging these values in to Equation~\ref{eq:christensen scaling law}, we estimate a dipolar field strength of $341_{-221}^{+239}$~G for J0623. Note that for simplicity, we sample the coefficient at the front of Equation~\ref{eq:christensen scaling law} from a Gaussian centered at 3.4~kG with a standard deviation of 2.2~kG, as we do not have access to the distribution of its value. 

Comparing the estimated dipolar field strength of J0623 to the minimum values estimated in Section~\ref{sec:field strength}, the value of 341~G is significantly lower. This same result was previously reported by \citet{kao18} for other T dwarfs. One way to reconcile this discrepancy is that the emission is in fact coming from small loops associated active regions \citep[e.g.][]{lynch15}. However, modelling each AFL with its own magnetic axis introduces more free parameters, which we do not account for in this work. Additionally, the ability for brown dwarfs to sustain such active regions over a significant length of time is uncertain, as pointed out by \citet{kao18}. Alternatively, it may also occur at the harmonic of the local cyclotron frequency, which would half the field strength estimate. One thing to note here is that our method provides a larger lower limit compared to previous works, as it is informed by the magnetic co-latitude of the emission cone. This may cause further deviations from the scaling law proposed by \citet{christensen09} if applied to other radio-detected UCDs.



\section{Potential plasma sources for J0623}
\label{sec:plasma}

The source of plasma powering both coherent and incoherent radio emission from the magnetospheres of isolated UCDs remains a major open question. One option is that it is driven by a sub-stellar wind. \citet{leto21} proposed a mechanism dubbed `centrifugal breakout' (CBO), in which material driven by a wind outflow builds up in the equatorial region around the UCD. The centrifugal force pushes this material outwards, but it is held in place by the strong magnetic field. Eventually, the equator reaches a critical mass at which the magnetic field can no longer hold on to the material, and it is flung outwards, snapping open closed field lines in the process. These subsequently reconnect, releasing energy and driving electron acceleration towards the magnetic poles.

\citet{leto21} illustrated that the CBO mechanism can reproduce the observed radio luminosities of a few late M-dwarfs and early L-dwarfs. However, it seems less promising for cooler UCDs, given that they do not exhibit any signs of coronae \citep{williams14, magaudda24}, which is a key ingredient for driving a wind outflow at low masses \citep[although see the recent work by][]{walters23}. Alternatively, orbiting satellites may feed plasma to the magnetosphere via outgassing, analogous to Jupiter's moon Io \citep{neubauer80}. Similar to the CBO mechanism, this material may build up within the equatorial region, which could cause reconnection analogous as described in Section~\ref{sec:MeerKAT fitting}. 

The co-rotation radius depends on the mass and rotation rate \citep{jardine19}:
\begin{equation}
R_\text{c} = \bigg(\frac{GMP^2}{4\pi^2}\bigg)^{1/3} ,
\end{equation}
where $P$ is the rotation period. Using the mass and radius estimates for J0623 from Section~\ref{sec:field strength}, we estimate a small co-rotation radius of $2.07_{-0.58}^{+0.79}~R$. This is unsurprising given its rapid rotation. It is interesting that the co-rotation radius is smaller than the minimum loop size inferred at the 3 epochs where radio emission was observed (Section~\ref{sec:field strength}). This could resemble a scenario similar to CBO, in that the AFLs would interact with plasma outside of the co-rotation radius. However, it is unclear as to if the timescale for CBO corresponds to the `lifetime' of the bursts seen on J0623. Nevertheless, it is also worth noting that the concept of material forced into co-rotation via the magnetic field is reminiscent of the recent findings by \citet{bouma24} indicative of co-rotating clumps of gas around pre-main sequence M dwarfs (referred to as complex periodic variables).

The question also still remains of why only two specific field lines emit within the magnetosphere of J0623. It could be that the field has a shape that is more complex than a dipole, and these lines are simply those that are visible to us. Alternatively, small loops associated with active regions could mimic the signature we infer with the 2 AFLs \citep{lynch15}. However, given the stability of the lightcurve between Dec 2022 and Mar 2023 (Figure~\ref{fig:lightcurve comp}), this is difficult to reconcile.


\section{Conclusions}

In this work, we developed a new tomographic technique for extracting the magnetospheric properties of planetary-mass objects from observations of coherent radio bursts. For ultracool dwarfs (UCDs), this information can provide a wealth of information pertaining to their internal composition and dynamo generation. With dedicated monitoring of radio-loud systems such as J0623 with telescopes like MeerKAT, we may also be able to unveil magnetic cycles in action on UCDs for the first time.

It is worth noting that our method could inform signatures of `auroral'-like emission on UCDs at other wavelengths. For instance, \citet{hallinan15} showed that the M8.5 dwarf LSR J1835+3259 exhibits modulated H$\alpha$ over the same timescale as its coherent radio bursts. This H$\alpha$ emission therefore may also be encoded with information about the magnetic field geometry, which could inform our radio modelling and visa versa. Additionally, JWST has begun to uncover emission features indicative of auroral processes on UCDs \citep[e.g.][]{faherty24}. These examples highlight the benefit of multi-wavelength observations

The viewing geometry and magnetic field configuration we infer for J0623 are of particular note, in that they co-incide with the same geometry identified by \citet{kavanagh23} which radio surveys may be biased towards. Therefore, other radio-detected UCDs may also be in a similar configuration. In terms of biases, \citet{kao24} also recently demonstrated that detection of quiescent radio emission from UCDs is more favourable in binaries. This could also be due to geometric effects, and/or enhanced energetics powering the emission. Whether this also holds for coherent radio emission from UCDs remains to be seen, however our findings may inform this. 

In terms of our model, an important aspect it does not capture is propagation effects. Depending on the plasma conditions in the magnetosphere, ECM emission may refract and or be absorbed as it propagates outwards. This naturally will alter the visibility of the emission, which in turn will affect the inferred geometry. However, developing a model that captures these processes which can interface with sampling algorithms efficiently is likely a complex task. Additionally, we currently have little-to-no information about the plasma environments surrounding UCDs. Therefore, these efforts will require additional parameterisation, further increasing the dimensionality of the problem. We also note there are hints of additional sub-structure within the lightcurves of J0623 that our model does not capture. These could be due to a field that is more complex than our assumed dipolar field shape, which could include smaller-scale surface features as have been inferred on other UCDs \citep{lynch15}. Accounting for more complex fields will also increase the dimensionality of the problem. However, this is worth exploring further from an inversion perspective.


\begin{acknowledgements}

RDK, HKV, and SB acknowledge funding from the Dutch Research Council (NWO) for the project `e-MAPS' (project number Vi.Vidi.203.093) under the NWO talent scheme VIDI. HKV also acknowledges funding from the European Research Council for the project `Stormchaser' (grant number 101042416). KR acknowledges the LSST-DA Data Science Fellowship Program, which is funded by LSST-DA, the Brinson Foundation, and the Moore Foundation; Their participation in the program has benefited this work. The MeerKAT telescope is operated by the South African Radio Astronomy Observatory, which is a facility of the National Research Foundation, an agency of the Department of Science and Innovation. The Australia Telescope Compact Array is part of the Australia Telescope National Facility, which is funded by the Australian Government for operation as a National Facility managed by CSIRO. We acknowledge the Gomeroi people as the Traditional Owners of the Observatory site. We thank Dr Zhoujian Zhang for providing their sampling chains from \citet{zhang21} for the effective temperature and surface gravity of J0623, which were used to estimate its mass, radius, and luminosity. We also thank Dr Joe Callingham for sharing their comments and suggestions on the manuscript.

\textit{Software}: NumPy \citep{numpy}, SciPy \citep{scipy}, pandas \citep{pandas}, Matplotlib \citep{matplotlib}, corner \citep{corner}, UltraNest \citep{buchner16, buchner19, buchner21a}.

\end{acknowledgements}


\bibliographystyle{aa}
\bibliography{bibliography}


\begin{appendix}
\section{ECM emission from active field lines}
\subsection{Precession of the magnetic axis}
\label{sec:coordinates}

The visibility of ECM emission as a function of time depends on the orientation of the magnetic axis $\boldhat{z}_\text{B}$ relative to the line of sight and the longitude of the AFL $\phi_\text{B}$. It is convenient to define the coordinates relative to the line of sight, which we do as follows. The line of sight $\boldhat{x} = (1,\,0,\,0)$ points from the center of the object to the observer. Note that vectors denoted with a hat are unit vectors. The rotation axis $\boldhat{z}_\text{rot}$ is inclined relative to $\boldhat{x}$ by the angle $i$: 
\begin{equation}
\boldhat{z}_\text{rot} = \cos i \boldhat{x} + \sin i \boldhat{z},
\end{equation}
where $\boldhat{z} = (0,\,0,\,1)$ is the projection of $\boldhat{z}_\text{rot}$ on to the plane of the sky. $\boldhat{y} = \boldhat{z}\times\boldhat{x} = (0,\,1,\,0)$ then completes the vector basis for the coordinate system.

We trace the rotation with the vector $\boldhat{x}_\text{rot}$. First, we project $\boldhat{x}$ onto the equatorial plane, which is
\begin{equation}
\boldhat{n}_\text{rot} = \sin i \boldhat{x} - \cos i \boldhat{z}.
\end{equation}
$\boldhat{x}_\text{rot}$ is measured from this vector, with which it forms the rotation phase angle
\begin{equation}
\phi_\text{rot} = \phi_0 + \frac{2\pi t}{P} = \phi_0 + \phi,
\end{equation}
i.e.
\begin{equation}
\boldhat{x}_\text{rot} = \sin\phi_\text{rot} \boldhat{y} + \cos\phi_\text{rot} \boldhat{n}_\text{rot} .
\end{equation}
In other words, the rotation phase is zero when $\boldhat{x}_\text{rot} = \boldhat{n}_\text{rot}$. Additionally, $\phi_0$ is the rotation phase at $t = 0$. Note that we also require a third vector here to complete the basis describing the rotation:
\begin{equation}
\boldhat{y}_\text{rot} = \boldhat{z}_\text{rot} \times \boldhat{x}_\text{rot} = \cos\phi_\text{rot} \boldhat{y} - \sin\phi_\text{rot} \boldhat{n}_\text{rot} .
\end{equation}

The plane defined by $\boldhat{z}_\text{rot}$ and $\boldhat{x}_\text{rot}$ contains the magnetic axis $\boldhat{z}_\text{B}$, which is inclined relative to $\boldhat{z}_\text{rot}$ by the magnetic obliquity $\beta$:
\begin{equation}
\boldhat{z}_\text{B} = \sin\beta\boldhat{x}_\text{rot} + \cos\beta\boldhat{z}_\text{rot}.
\end{equation}
When $\beta \neq 0$ or $180\degr$, the observer will see $\boldhat{z}_\text{B}$ precess about the rotation axis. It is also convenient to project $\boldhat{x}_\text{rot}$ on to the magnetic equator:
\begin{equation}
\boldhat{n}_\text{B} = \sin \beta \boldhat{x}_\text{rot} - \cos \beta \boldhat{z}_\text{rot}.
\end{equation}
This provides a reference vector to measure the magnetic longitude $\phi_\text{B}$ from for each AFL:
\begin{equation}
\boldhat{x}_\text{B} = \sin\phi_\text{B}\boldhat{y}_\text{rot} + \cos\phi_\text{B}\boldhat{n}_\text{B} ,
\end{equation}
i.e. this vector locates the AFL in the magnetic equator.


\subsection{Beam angle}
\label{sec:beam angle}

The plane formed by the vectors $\boldhat{z}_\text{B}$ and $\boldhat{x}_\text{B}$ contains the AFL, which is dipolar in shape. The unit vector $\boldhat{c}$ at each point on the field line, which is the normalised magnetic field vector, is \citep{bloot24}:
\begin{equation}
\boldhat{c} = a \boldhat{x}_\text{B} \pm b \boldhat{z}_\text{B},
\label{eq:cone vector}
\end{equation}
where $\theta_\text{B}$ is the magnetic co-latitude of the point measured from $\boldhat{z}_\text{B}$, and
\begin{gather}
a = \frac{3\sin\theta_B\cos\theta_B}{(1 + 3\cos^2\theta_B)^{1/2}}, \label{eq:coefficient a}\\
b = \frac{3\cos^2\theta_B - 1}{(1 + 3\cos^2\theta_B)^{1/2}} .
\label{eq:coefficient b}
\end{gather}
An emission cone on the AFL will be aligned with the vector $\boldhat{c}$. At any frequency, cyclotron emission can occur at two points on the AFL, one in each magnetic hemisphere. The line is symmetric, meaning that emission at a given frequency and co-latitude $\theta_\text{B}$ also occurs at co-latitude $\pi - \theta_\text{B}$. Hence the $\pm$ symbol in Equation~\ref{eq:cone vector}, which refers to the Northern and Southern magnetic hemisphere.

The angle $\boldhat{c}$ makes with the line of sight determines if emission is visible to the observer. We obtain this by taking the dot product:
\begin{equation}
\cos\gamma = a \boldhat{x}_\text{B}\cdot\boldhat{x} \pm b \boldhat{z}_\text{B}\cdot\boldhat{x}.
\label{eq:beam angle}
\end{equation}
Expanding the dot product terms above gives the following:
\begin{gather}
\boldhat{x}_\text{B}\cdot\boldhat{x} = c \sin\phi_\text{rot} + d \cos\phi_\text{rot} + e, \\
\boldhat{z}_\text{B}\cdot\boldhat{x} = f \cos\phi_\text{rot} + g ,
\end{gather}
where
\begin{gather}
c = - \sin i\sin\phi_B, \\
d = \sin i\cos\beta\cos\phi_B, \\
e = - \cos i\sin\beta\cos\phi_B, \\
f = \sin i\sin\beta, \\
g = \cos i\cos\beta.
\end{gather}
Rewriting Equation~\ref{eq:beam angle}, we have
\begin{equation}
\cos\gamma = ac \sin\phi_\text{rot} + (ad \pm bf) \cos\phi_\text{rot} + (ae \pm bg) .
\end{equation}
Expanding the terms $\sin\phi_\text{rot}$ and $\cos\phi_\text{rot}$ then gives the following expressions for the beam angle from the Northern and Southern hemispheres:
\begin{gather}
\cos\gamma_N = T_1 \sin\phi + T_2 \cos\phi + T_3 \label{eq:beam angle N}, \\
\cos\gamma_S = T_4 \sin\phi + T_5 \cos\phi + T_6,
\label{eq:beam angle S}
\end{gather}
where
\begin{gather}
T_1 = ac\cos\phi_0 - (ad + bf)\sin\phi_0 \\
T_2 = ac\sin\phi_0 + (ad + bf)\cos\phi_0 \\
T_3 = ae + bg \\
T_4 = ac\cos\phi_0 - (ad - bf)\sin\phi_0 \\
T_5 = ac\sin\phi_0 + (ad - bf)\cos\phi_0 \\
T_6 = ae - bg .
\end{gather}
Note that the model lightcurve does not depend on the emission frequency (see Appendix~\ref{sec:model frequency}). Additionally, some of the parameters Equation~\ref{eq:beam angle} depends on are degenerate, which we describe in Appendix~\ref{sec:degeneracies}.


\subsection{Degeneracies}
\label{sec:degeneracies}

Given that the forward model presented here is predominantly geometric, it is important to identify degeneracies it contains to aid the convergence of the sampling algorithm utilised. The degeneracies we identify are due to symmetries in the functions used to model the beam angle from each hemisphere. For instance, if one replaces $i$ with $\pi - i$, $\beta$ with $\pi - \beta$, and $\phi_\text{B}$ with $\pi - \phi_\text{B}$, the functional forms of Equations~\ref{eq:beam angle N} and ~\ref{eq:beam angle S} are unchanged.

An additional symmetry can be found in the expressions for $a$ and $b$ (Equations~\ref{eq:coefficient a} and \ref{eq:coefficient b}). For $\theta_\text{B}\in[0,\pi/2]$, there are two values of $\theta_\text{B}$ that give the same value of $a$. These are given by
\begin{equation}
\cos^2\theta_B = \frac{1}{6}(3-a^2\pm\sqrt{9-10a^2+a^4}),
\label{eq:theta_B degeneracy}
\end{equation}
which lie either side of the maximum of $a$ which occurs at $\cos\theta_\text{B} = 1 / \sqrt{3}$. So, swapping $\theta_\text{B}$ for the other solution to Equation~\ref{eq:theta_B degeneracy} gives the same value for $a$. Additionally, $a^2 + b^2 = 1$, so the value of $b$ is inverted for the second value of $\theta_\text{B}$. If the value of $\theta_\text{B}$ is swapped in this case, the following transformations can be made to retain the function forms of Equations~\ref{eq:beam angle N} and ~\ref{eq:beam angle S}:
\begin{enumerate}
    \item $i\rightarrow\pi - i$, $\phi_0\rightarrow\pi+\phi_0$, $\phi_\text{B}\rightarrow\pi+\phi_\text{B}$
    \item $\beta\rightarrow\pi - \beta$, $\phi_0\rightarrow\pi+\phi_0$, $\phi_\text{B}\rightarrow 2\pi-\phi_\text{B}$
\end{enumerate}
In order to avoid these degeneracies, we limit the range of values for $i$ and $\beta$ from 0 to $90\degr$. In other words, our inferred values for $i$ and $\beta$ cannot be distinguished between their supplementary angles. Therefore, we cannot tell the direction of rotation of the pole we see, nor the polarity of the magnetic pole we see. The latter is further obscured by the fact that we do not know the magnetoionic mode under which the ECM emission is operating for J0623, which will flip the sign of the circularly-polarised flux density measured from a region of a given magnetic polarity \citep{kavanagh22}.


\section{Data reduction}
\subsection{MeerKAT}
\label{sec:MeerKAT data prep}

The MeerKAT data presented by \citet{rose23} shows slight differences in structure from 950 to 1150 MHz and 1300 to 1500 MHz. To account for this in our sampling method, we took the $64$\,second time-averaged data from \citet{rose23} and created two lightcurves averaged over the $950$--$1150$\,MHz and $1300$--$1500$\,MHz frequency sub-bands, respectively. The uncertainties for each flux density measurement are the standard error obtained by computing the standard deviation of the imaginary component of the emission across the frequency sub-band and normalising by the square root of the number of channels in that band. We then replace any missing values across the observation with the arithmetic mean of these standard error uncertainties.

We then phase folded the two lightcurves using the rotation period inferred by \citet{rose23} of 1.912 hours. From rotation to rotation, the structure in the lightcurves shows some variation. This is particularly noticeable in the first burst in the high frequency band \citep[see Figure~2 of][]{rose23}. However, our model assumes a fixed flux density for each AFL. To account for these intrinsic variations, we first define bins in phase space, each with a width equal to the time resolution of the MeerKAT data divided by the rotation phase. Then, we resample the points in each phase bin from a Gaussian distribution 10\,000 times. We then compute the average value and standard deviation of all the resampled points in each bin. As a result, the error in phase for each bin contains both the measurement error for each point and the intrinsic variations from rotation to rotation. With the data formatted, we can then compute the likelihood function for UltraNest to sample.


\subsection{ATCA}
\label{sec:ATCA data reduction}

J0623 was observed with the Australia Telescope Compact array twice in 2022 (Project ID: C3363, PI: T Murphy), at 1.1 to 3.1~GHz using the 2~GHz Compact Array Broadband Backend \citep[CABB,][]{wilson11}. The first observation was taken on 2022-08-14, in the H168 configuration, with an observing time of 6\,h. The second observation lasted for 11 hours and was taken in the 6D configuration.

We reduced the data using \texttt{casa} version 6.4 \citep{casa}. The data was flagged using the \texttt{tfcrop} algorithm, resulting in on average 50-60\% of the data being flagged. We use PKS\,B1934-638 as the flux density and bandpass calibrator, and J0607-157 as the phase calibrator. PKS\,B1934-638 was observed for 10 minutes at the start of each observation. The phase calibrator was observed for 2 minutes after every 20 minutes on target. 

We first determined the gain, bandpass, polarisation and polarisation leakage solutions on the flux density calibrator, using a solution interval of 60\,s. After transferring these solutions to the phase calibrator, we use the \texttt{qufromgain} task from the \texttt{atca.polhelpers} package for linear polarisation calibration. The combined solutions were then transferred to the target.

For both epochs, we create Stokes~V light curves of either 3 or 4 frequency bins as described in Section~\ref{sec:cycle}, using the method described in \citet{bloot24}. We define the sign of Stokes~V as right-handed circularly-polarised emission minus left-handed circular polarised emission, in agreement with the IAU convention \citep{hamaker96}. We then phase-folded and binned the data in the same manner as for the MeerKAT data.


\section{Parameter space exploration with UltraNest}
\label{sec:sampling}
\subsection{Computing the likelihood function}

The main component of running a sampling algorithm such as UltraNest is setting up the likelihood and posterior functions. In the case of the 2 MeerKAT bands, the lightcurves show little variation in frequency. To account for this in our likelihood function, we compute the model lightcurve $F_\nu$ for 8 frequencies spaced uniformly over each band via Equation~\ref{eq:line flux}. We then compute the total log likelihood by summing up the log likelihood of each model lightcurve with its respective band of observed emission $F_\text{obs}$:
\begin{equation}
\log\mathcal{L}_\nu = -\frac{1}{2} \sum \bigg\{\log(2\pi\sigma_\text{obs}^2) + \Big(\frac{F_\nu - F_\text{obs}}{\sigma_\text{obs}}\Big)^2 \bigg\},
\label{eq:likelihood}
\end{equation}
where $\sigma_\text{obs}$ is the standard deviation in each phase bin, which is computed as described in Appendix~\ref{sec:MeerKAT data prep}. Computing the model lightcurve $F_\nu$ requires the magnetic co-latitude of emission at each frequency (Appendix~\ref{sec:beam angle}). We describe our method for calculating this in Appendix~\ref{sec:model frequency}.


\subsection{The magnetic co-latitude of a given frequency}
\label{sec:model frequency}

Our lightcurve model does not depend on the emitted frequency (Appendix~\ref{sec:beam angle}). Instead, it depends on the magnetic co-latitude of the emitting point $\theta_\text{B}$. The magnetic field strength on a dipolar field line is \citep{kivelson95}:
\begin{equation}
B = \frac{B_0}{2} \Big(\frac{R}{r}\Big)^3 (1 + 3\cos^2\theta_\text{B})^{1/2} ,
\end{equation}
where $B_0$ is the field strength at the magnetic poles, $R$ is the radius, and $r$ is the radial distance to the emitting point from the center. The coordinates on a dipolar field line also relate to one another via:
\begin{equation}
r = L\sin^2\theta_\text{B},
\label{eq:field coordinates}
\end{equation}
where $L$ is the loop size measured from the center to its maximum extent in the magnetic equator. Additionally, fundamental cyclotron emission occurs at a frequency of $\nu = 2.8~B$~MHz, where $B$ is in Gauss. Combining all this together, one can show that
\begin{equation}
1.4 B_0 \Big(\frac{R}{L}\Big)^3 = \nu \frac{\sin^6\theta_\text{B}}{(1+3\cos^2\theta_\text{B})^{1/2}} .
\label{eq:constant}
\end{equation}
Our sampling routine provides us with the value of $\theta_\text{B}$ that best-reproduces the observed lightcurve. Assigning the frequency of emission $\nu$ at this magnetic co-latitude therefore means that the left-hand side of Equation~\ref{eq:constant} is constant. In other words, one can increase both the dipolar field strength and loop size without changing the lightcurve morphology.

In our sampling routine, we first draw the co-latitude $\theta_\text{max}$ corresponding to the maximum frequency $\nu_\text{max}$. Since Equation~\ref{eq:constant} is constant for each AFL, we can write an expression relating the co-latitude at $\nu_\text{max}$ to the co-latitude $\theta_\text{B}$ corresponding to emission at any frequency $\nu$:
\begin{equation}
\nu_\text{max} \frac{\sin^6\theta_\text{max}}{(1+3\cos^2\theta_\text{max})^{1/2}} = \nu \frac{\sin^6\theta_\text{B}}{(1+3\cos^2\theta_\text{B})^{1/2}} .
\label{eq:freq comp}
\end{equation}
Re-writing, we have:
\begin{equation}
\frac{x^6}{4 - 3x} = Q,
\label{eq:x}
\end{equation}
where
\begin{gather}
Q = \Big(
\frac{\nu_\text{max}}{\nu}\Big)^2 \frac{{x_\text{max}}^6}{4 - 3x_\text{max}} , \\
x = \sin^2\theta_\text{B} , \\
x_\text{max} = \sin^2\theta_\text{max} .
\end{gather}
To solve Equation~\ref{eq:x}, we use Newton's method, i.e.:
\begin{equation}
x_{i+1} = x_i - \frac{{x_i}^6 + 3Qx_i - 4Q}{6{x_i}^5 + 3Q}
\label{eq:newton}
\end{equation}
Initialising the value of $x_i = x_\text{max}$, the relative difference between the left and right-hand side of Equation~\ref{eq:freq comp} becomes less than $10^{-10}$ within 5 iterations. We solve Equation~\ref{eq:newton} for each of the 8 uniformly-spaced frequencies over the observed band. Note that this is done independently for each AFL. With the co-latitude at each frequency obtained, we can then compute the likelihood via Equation~\ref{eq:likelihood}.


\subsection{Adopted priors for the parameter space}

We also must specify the priors each model parameter. The full list of parameters is as follows: 
\begin{itemize}
\item The inclination of the rotation axis $i$
\item The magnetic obliquity $\beta$
\item The rotation phase at the start of the MeerKAT observations $\phi_0$
\item The emission cone opening angle $\alpha$ and thickness $\Delta\alpha$
\item The magnetic co-latitude of the emission site on each AFL ($\theta_1, \theta_2, ...$)
\item The longitude of the AFL in the magnetic equator ($\phi_1, \phi_2, ...$)
\item The flux density measured from each AFL when the beam direction is aligned exactly with the line of sight ($F_1, F_2, ...)$
\end{itemize}
We have little information about any of these quantities, so we adopt uniform priors for each. Note that for the inclination $i$, we uniformly sample $\cos i$ \citep[see][]{kavanagh23}. The inclination and magnetic obliquity are also constrained by the degeneracies as described in Appendix~\ref{sec:degeneracies}. On Jupiter, opening angles of around 70 to 80$\degr$ and thicknesses of a few degrees have been inferred from in-situ observations \citep{kaiser00, louis23}, although lower opening angles have also been inferred from theoretical models \citep{stupp00}. Given the uncertainties and complexities leading to the the resulting cone parameters of ECM emission, we choose to vary the opening angle over 90 degrees, and consider thicknesses up to 10 degrees. 

Another constraint to note is on the max value of the co-latitude of the emission cone. In our sampling routine, our input value is the co-latitude at the maximum frequency $\nu_\text{max}$. However, if the co-latitude is too large, it will not be possible to generate emission on the AFL at the lowest frequency $\nu_\text{min}$. To avoid this, we limit the value of $\theta_\text{B}$ at $\nu_\text{max}$ to be less the value which places emission at $\nu_\text{min}$. Re-writing Equation~\ref{eq:freq comp}, we have
\begin{equation}
\nu_\text{max} \frac{\sin^6\theta_\text{max}}{(1+3\cos^2\theta_\text{max})^{1/2}} = \nu_\text{min} \frac{\sin^6\theta_\text{min}}{(1+3\cos^2\theta_\text{min})^{1/2}} .
\label{eq:theta max}
\end{equation}
where $\nu_\text{min} = 950$~MHz and $\nu_\text{max} = 1500$~MHz for the MeerKAT data, and $\nu_\text{min} = 1.3$~GHz and $\nu_\text{max} = 3.1$~GHz for the ATCA data. Setting $\theta_\text{min} = 90\degr$, we solve Equation~\ref{eq:theta max} as described in Appendix~\ref{sec:model frequency}. For the MeerKAT data, we obtain $\theta_\text{max} = 71.36\degr$, and for the ATCA data $\theta_\text{max} = 63.98\degr$.

For the flux density of each AFL, we set the minimum value to be 1~mJy, which is the lowest burst flux seen in the MeerKAT data. The upper limit of 10~mJy is chosen to allow the sampler to also explore scenarios where significant flux cancellation occurs between emission cones in the two magnetic hemispheres. The remaining parameters are self explanatory, i.e. the initial phase and magnetic longitude of each AFL are uniformly sampled in on a circle. We also note that we enforce the value of $\phi_B$ for AFL~2 to be greater than that for AFL~1. Table~\ref{table:prior ranges} lists all of the priors.

\begin{table}
\caption{List of the priors adopted for each parameter in our model. Each value is drawn from a uniform distribution $\mathcal{U}$, with the proceeding numbers indicating the lower and upper limit. Throughout the paper, the subscript `B' is interchanged for the AFL number (in our case, 1 or 2).}
\label{table:prior ranges}
\centering
\begin{tabular}{cc}
Parameter & Value \\
\hline
$\cos i$ & $\mathcal{U}(0, 1)$ \\
$\beta$ ($\degr$) & $\mathcal{U}(0, 90)$ \\
$\phi_0$ & $\mathcal{U}(0, 1)$ \\
$\alpha$ ($\degr$) & $\mathcal{U}(0, 90)$ \\
$\Delta\alpha$ ($\degr$) & $\mathcal{U}(0, 10)$ \\
$\theta_\text{B}$ ($\degr$) & $\mathcal{U}(0, \theta_\text{max})$ \\
$\phi_\text{B}$ ($\degr$) & $\mathcal{U}(0, 360)$ \\
$F_\text{B}$ (mJy) & $\mathcal{U}(1, 10)$ \\
\end{tabular}
\end{table}


\subsection{Running the sampler}

\begin{figure*}
\centering
\includegraphics[width=\textwidth]{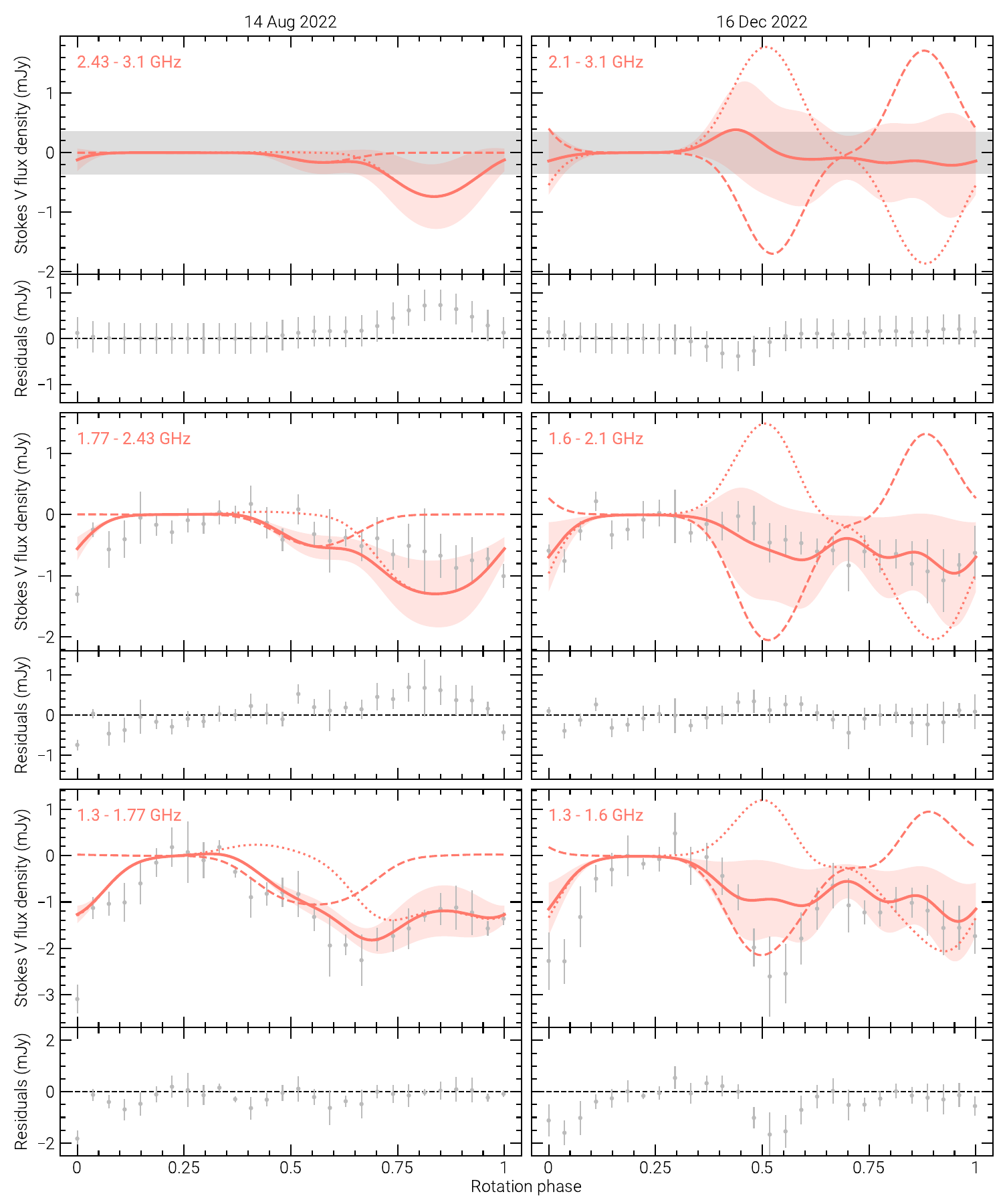}
\caption{Same as Figure~\ref{fig:MeerKAT lightcurves} except for the best fit to the Aug and Dec ATCA data. The parameters inferred at these two epochs are listed in Table~\ref{table:ATCA values}. The top panels show the $1\sigma$ `noise' bands in which we do not find any significant emission. Note that the rotation phases shown here are consistent with those in Figure~\ref{fig:MeerKAT lightcurves}.}
\label{fig:ATCA lightcurves}
\end{figure*}

\begin{figure*}
\includegraphics[width = \textwidth]{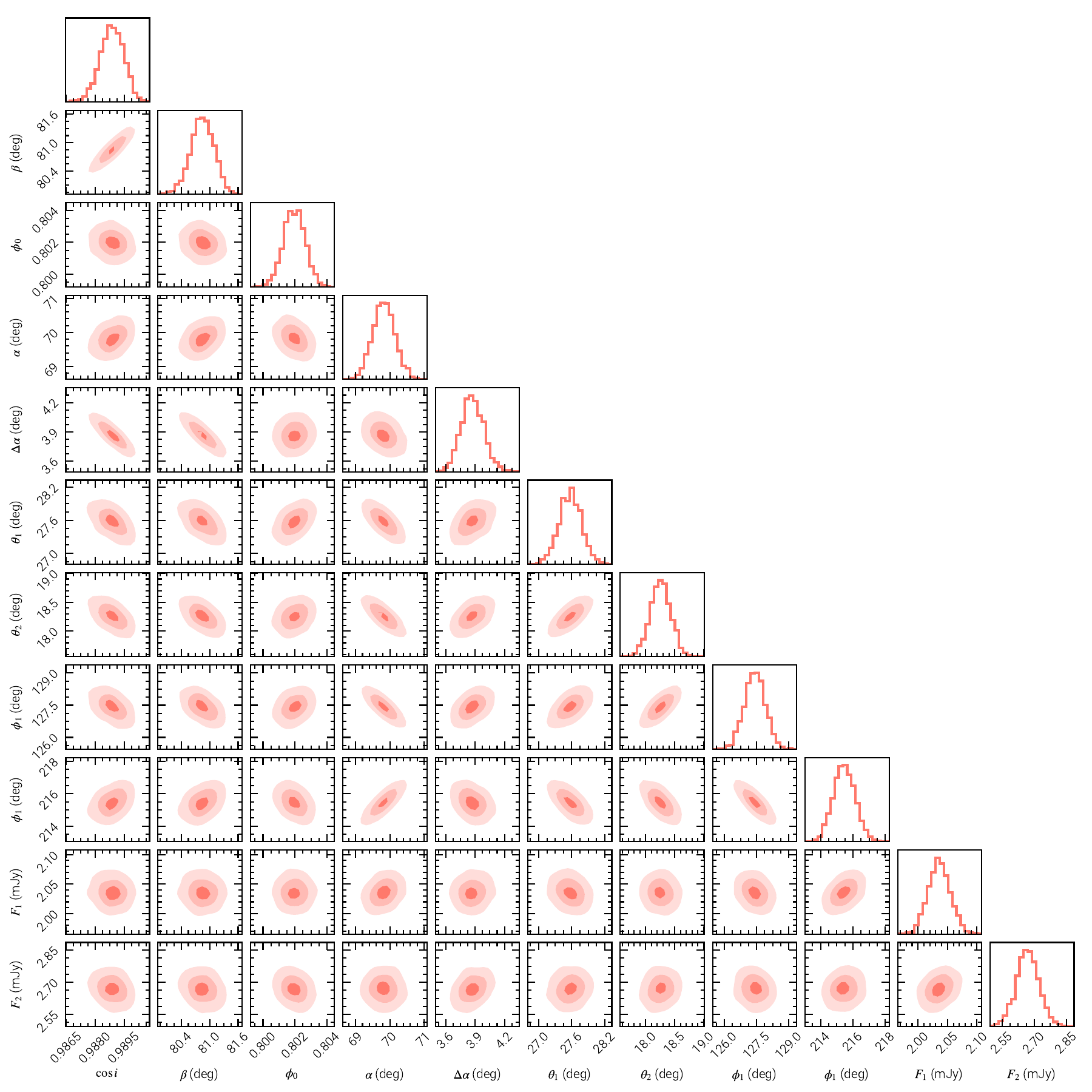}
\caption{Posterior distributions of the best-fitting model parameters listed in Table~\ref{table:MeerKAT values} plotted using \texttt{corner} \citep{corner}. The shaded regions in each panel bar the diagonal ones show the 16th, 50th, and 84th percentiles going from darkest to lightest in colour. Note that we have smoothed the contours by setting the \texttt{smooth} parameter to 1 in the \texttt{corner} function to better-highlight the 16th percentile region. The diagonal panels show the flattened histogram for each parameter (unsmoothed).}
\label{fig:MeerKAT corner plot}
\end{figure*}

\begin{figure*}
\centering
\includegraphics[width=0.495\textwidth]{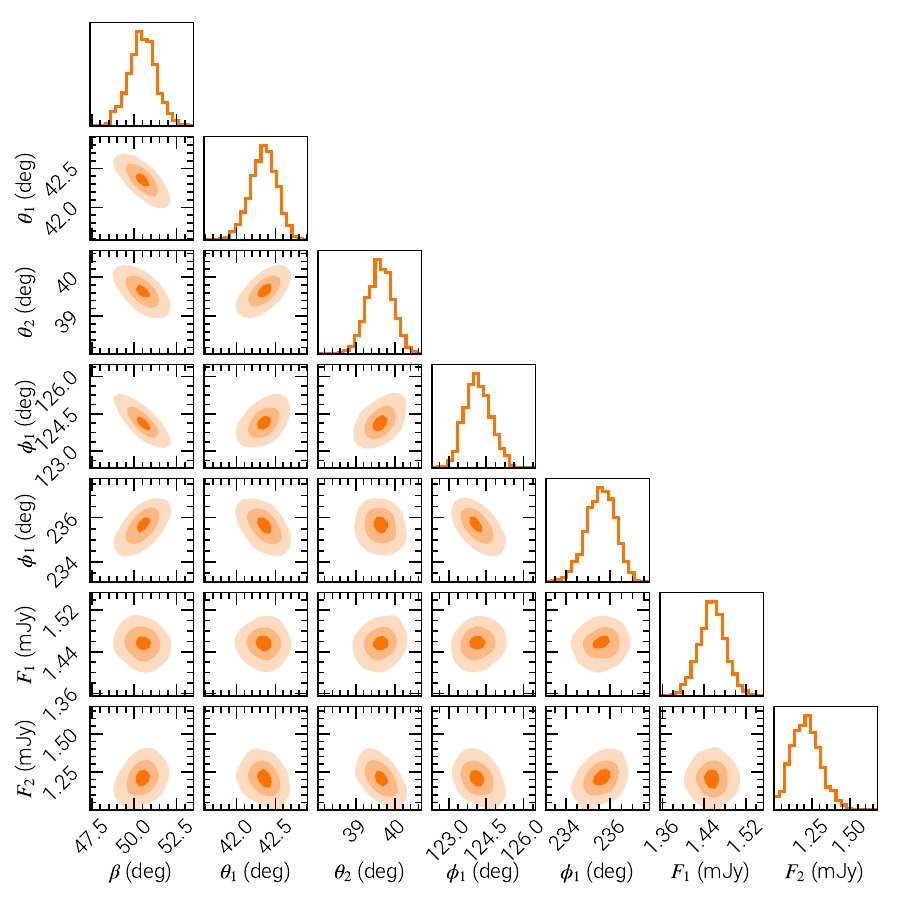}
\includegraphics[width=0.495\textwidth]{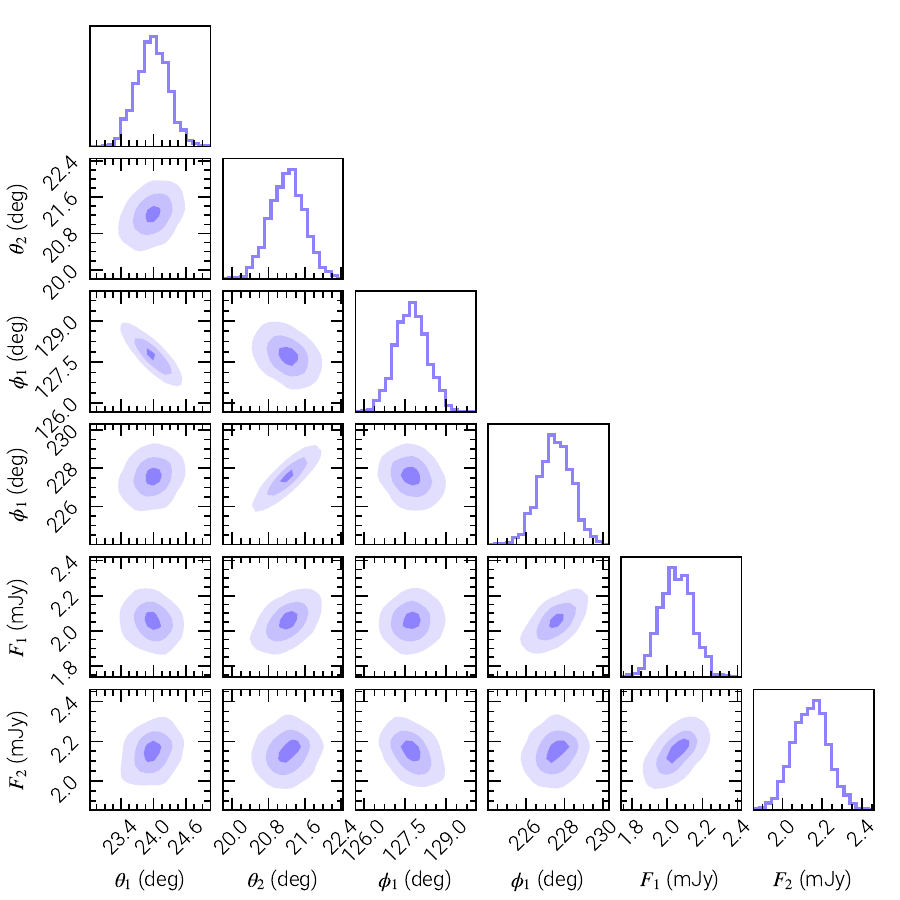}
\caption{Same as Figure~\ref{fig:MeerKAT corner plot} except for the Aug (left) and Dec (right) ATCA data. The derived confidence intervals for each parameter are listed in Table~\ref{table:ATCA values}. Dec data favours the magnetic magnetic obliquity $\beta$ to be fixed as the value inferred from the MeerKAT data.}
\label{fig:ATCA corner plots}
\end{figure*}

With the likelihood and prior functions set up, we then use UltraNest's step sampler function \texttt{SliceSampler}, which is recommended for efficient sampling in high dimensions\footnote{\url{https://johannesbuchner.github.io/UltraNest/example-sine-highd.html}}. For the \texttt{generate\_direction} parameter of \texttt{SliceSampler}, we use \texttt{generate\_mixture\_random\_direction}, which performs best in high dimensional spaces \citep{buchner21b}. For each dataset and number of AFLs, we run the sampler multiple times and manually check if it converges on the same result. If the results are inconsistent, we double the number of steps the sampler takes via the \texttt{nsteps} parameter. For the MeerKAT data, the sampler converges in 200 steps for 1 AFL and 400 steps for 2 AFLs, whereas for the ATCA data it converges in 100 steps for both 1 and 2 AFLs. Our best fits to the MeerKAT and ATCA data are shown in Figures~\ref{fig:MeerKAT lightcurves} and \ref{fig:ATCA lightcurves}. We also show the converged posterior distributions for the MeerKAT and ATCA data in Figures~\ref{fig:MeerKAT corner plot} and \ref{fig:ATCA corner plots}. An example script showing our setup for sampling the likelihood space defined by our model can be found on GitHub\footnote{\url{https://github.com/robkavanagh/J0623}}.

\end{appendix}
\end{document}